# Unconventional co-existence of insulating nano-regions and conducting filaments in reduced $SrTiO_3$: mode softening, local piezoelectricity and metallicity

Annette Bussmann-Holder[1], Hugo Keller[2], Arndt Simon[1], Krystian Roleder[3], and Krzysztof Szot[3,4]

[1]Max-Planck-Institute for Solid State Research, Heisenbergstr. 1, D-70569 Stuttgart, Germany

[2]Physik-Institut der Universität Zürich, Winterthurerstr. 190, Ch-8057 Zürich, Switzerland

[3]Institute of Physics, University of Silesia, ul. 75 Pułku Piechoty 1, 41-500 Chorzów, Poland

[4]Institute of Solid State Research, Forschungszentrum Jülich, D-52425 Jülich, Germany

Correspondence and requests for materials

should be addressed to A. B.-H. (email: a.bussmann-holder@fkf.mpg.de)

***Abstract***

Doped $SrTiO_3$ becomes a metal at extremely low doping concentrations $n$ and is even superconducting at $n < 10^{20}cm^{-3}$ with the superconducting transition temperature adopting a dome-like shape with the carrier concentration. It is shown here within the polarizability model that up to a well-defined carrier concentration $n_c$ transverse optic mode softening takes place together with polar nano-domain formation which evidences inhomogeneity and a two-component type behavior with metallicity coexisting with polarity. Beyond this region a conventional metal is formed where superconductivity as well as mode softening is absent. For $n \leq n_c$ the effective electron-phonon coupling follows the superconducting transition temperature. Effusion measurements as well as macroscopic and nanoscopic conductivity measurements indicate that the distribution of oxygen vacancies is local and inhomogeneous from which it is concluded that metallicity stems from filaments which are embedded in a polar matrix as long as $n \leq n_c$.

***Introduction***

$SrTiO_3$ (STO) is one of the best investigated systems for scientific and technological reasons. Early on the research activities focused on the cubic to tetragonal phase transition at $T_S$=105 K [1 – 6] and the transverse optic mode softening suggests a polar instability [7 – 10]. While the structural instability was early assigned to the rotation of oxygen octahedron driven by a soft transverse acoustic zone boundary mode, the polar instability has been shown to be suppressed by quantum fluctuations and the phase named quantum paraelectric and/or incipient ferroelectric, respectively [1, 5]. Ferroelectricity can, however, be induced by either isotopic substitution of $^{16}O$ by $^{18}O$ [11], or by replacing tiny amounts of Sr by Ca [12]. In the former case the polar state remains incomplete [13, 14] since it is inhomogeneous with coexisting paraelectric and ferroelectric domains, whereas in the latter an XY pseudospin ferroelectric transition takes place [12]. Recently, it has been shown that the domain walls of STO in the tetragonal phase carry polar properties related to ferroelectricity [15, 16]. Upon replacing Sr by Ca or La, or Ti with Nb, or removing oxygen, superconductivity has been observed at low temperature with a dome like dependence of the transition temperature upon doping [17 – 25]. In all cases the carrier concentrations are extremely small and the Fermi energies an order of magnitude smaller than the Debye temperature. For oxygen deficient samples superconductivity has early on been explained as soft mode [26 – 28] or intervalley deformation potential driven [29, 30, 23]. Another interpretation has combined both proposals by introducing unconventional anharmonicity [31]. Recent novel interest in STO has invoked ferroelectric mode fluctuations [32], longitudinal optical phonons [33] or plasmons [34] as possible source of superconductivity. Besides of its superconducting properties also the metallic state of STO and its metal / insulator (M/I) transition have attracted intense attention, since metallicity is observed at a carrier concentration which reaches a critical value defined by the Mott's criterion [35]. This transition can be induced through reduction in high temperature and under low oxygen partial pressure, e.g., in vacuum, Ti-getter or $H_2$ atmosphere [18, 36]. *Macroscopic* measurements of the Hall effect indeed show that the critical concentration of free carriers required for the M/I transition should be of the order of $n \approx 10^{16}/cm^3$ [25, 36]. However, effusion measurements and macroscopic and nanoscopic conductivity measurements show that the removal of only $10^{13}$-$10^{14}$ oxygen ions/$cm^3$ turns STO into a metallic state [37, 38] and thus violate the Mott criterion. In addition, such an extremely low distribution of oxygen vacancies gives rise to local inhomogeneity which implies that the reduction process is of local character and takes place only near the core of dislocations [37 – 39]. This so called self-doping is limited to the network of dislocations, increasing the *local* concentration of Ti $d^1$ electrons to $10^{20}$-$10^{21}/cm^3$ [37, 40, 41]. These observations imply that the Mott criterion is satisfied *locally,* whereas from the point of view of average properties it is violated.

In this paper the dynamical properties of doped STO are investigated and shown to be related to superconductivity. Especially, it is demonstrated that the soft optic mode persists up to a critical carrier concentration $n_c$ and leads to the formation of polar nano domains. Since ferroelectricity and also incipient ferroelectricity as realized in undoped STO are incompatible with metallicity, an inhomogeneous state emerges where the metallic conductivity is filamentary [39, 42, 43], whereas the mode softening takes place in the embedding matrix which consists of polar nanoregions. Interestingly, superconductivity is confined to the regime of coexistence, thus highlighting the importance of the pseudo ferroelectric soft phonon mode. These results are supported by an experimental conductivity study which evidences the filamentary character of the conductivity in coexistence with polar nano domains.

### *Theoretical background*

The dynamical properties of STO have been studied comprehensively experimentally since its discovery [2 – 4, 6, 9, 10]. Theoretical intense work has been devoted to their understanding based on phenomenological modelling as well as *ab initio* theories. Especially the presence of strong anharmonicity has been emphasized when concentrating on the essential temperature dependent properties of this compound. A rather profound breakthrough was obtained by Migoni et al. [44], who introduced the nonlinear polarizability model to successfully reproduce quantitatively the characteristic dynamical properties of ferroelectric perovskites including STO within the self-consistent phonon approximation (SPA). The model has been cast into a more transparent pseudo-one-dimensional version and applied to a broad range of ferroelectrics [45 – 48]. The essential ingredient of the model as introduced by Migoni et al [44] is based on the nonlinear polarizability of the oxygen $O^{2-}$ ion which depends strongly on its crystalline environment. In Ref. 44 this was accounted for by using two independent core-shell force constants with respect to the A and B neighbors in $ABO_3$. The lowest term of the nonlinear core-shell interaction at the oxygen-ion site is of fourth order because of inversion symmetry in the paraelectric phase. However, the temperature dependence of the soft mode and other low frequency modes as well as the Raman spectra depends on the coupling constant directed to the transition metal ion only. This corresponds to the modulation of the oxygen-ion polarizability in the direction of the neighboring Ti ion in STO and indicates that the phonon-induced change of the transition-metal —oxygen bond, i.e., the hybridization of oxygen p and transition metal d electrons, plays an essential role for the dynamical properties of ferroelectrics ("dynamical covalency"). The importance of the p-d hybridization has been demonstrated for various perovskites and successfully interpreted on the basis of nonlinear polarizability model. The onsite potential in the core–shell interaction is of double-well character with $g_2$ being the attractive harmonic coupling constant and $g_4$ the fourth order repulsive anharmonic coupling constant, and can be written as $V(w_{1n}^2) = \frac{1}{2} g_2 w_{1n}^2 + \frac{1}{4} g_4 w_{1n}^4$ using the definition $w_{1n} = u_{1n} - v_{1n}$ where $u_{1n}$ is the ionic displacement and $v_{1n}$ the electronic shell related one, respectively, at lattice site *1n*. The relative displacement coordinate characterizes the polarization and has the limits of a fully delocalized shell for $w_{1n} \to \infty$, i.e. complete ionization, and a rigidly bound shell for $w_{1n} = 0$ where p-d hybridization effects are irrelevant. In metallic samples new or in-gap states appear at or close to the Fermi energy which diminish these effects and correspondingly destroy the proximity to a polar instability.

A review on the model is given in Ref. [45, 46]. For doped semiconducting STO the observed soft transverse optic zone center mode has been shown to harden with increasing carrier concentration [49] and a linear relation between carrier concentration and $g_2$ as well as $g_4$ was established [50].

In the following this relation is used to extract $g_2$ as well as $g_4$ for arbitrary carrier concentrations. This method allows us to obtain the related double-well potential and to calculate within the SPA the optic mode softening, coupling of optic and acoustic mode, spatial extensions of polar nano-domains and effective electron-phonon coupling constant as a function of carrier concentration *n*.

In Figure 1 $g_2$ and $g_4$ are shown as a function of *n*. Obviously the nonlinear interaction constant $g_4$ is small and almost independent of *n*, whereas $g_2$ rapidly increases with increasing *n* and changes sign for $n_c$=2.232×$10^{20}$ $cm^{-3}$, the carrier concentration for which the superconducting transition temperature $T_c$ vanishes.

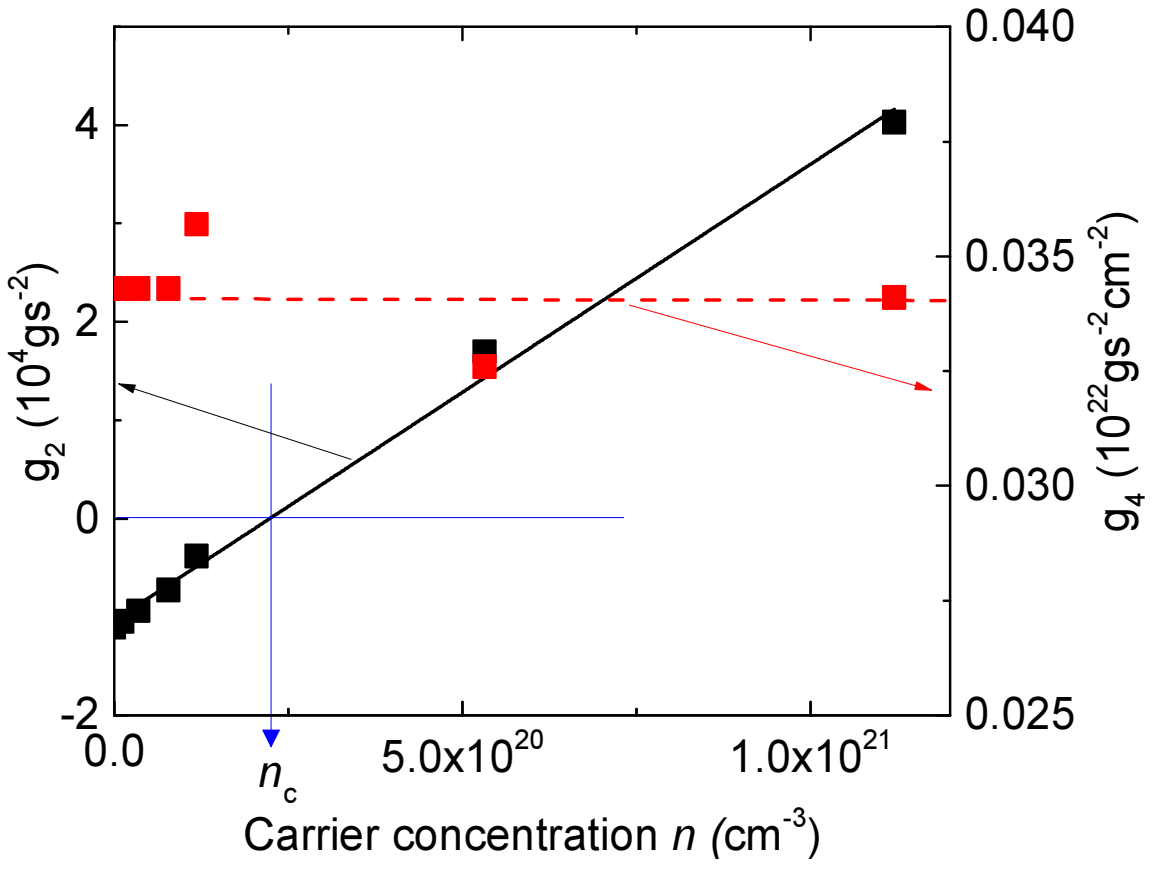

**Figure 1** Carrier concentration dependence of the harmonic $g_2$ and the fourth order anharmonic $g_4$ coupling constants. Note, that $g_2$ changes sign at $n_c$ from attractive to repulsive.

At $n_c$ the double-well potential changes to a single-well one (Figure 2) where the depicted carrier concentrations refer to the ones discussed in [51, 52] in order to provide the connection to experimental data.

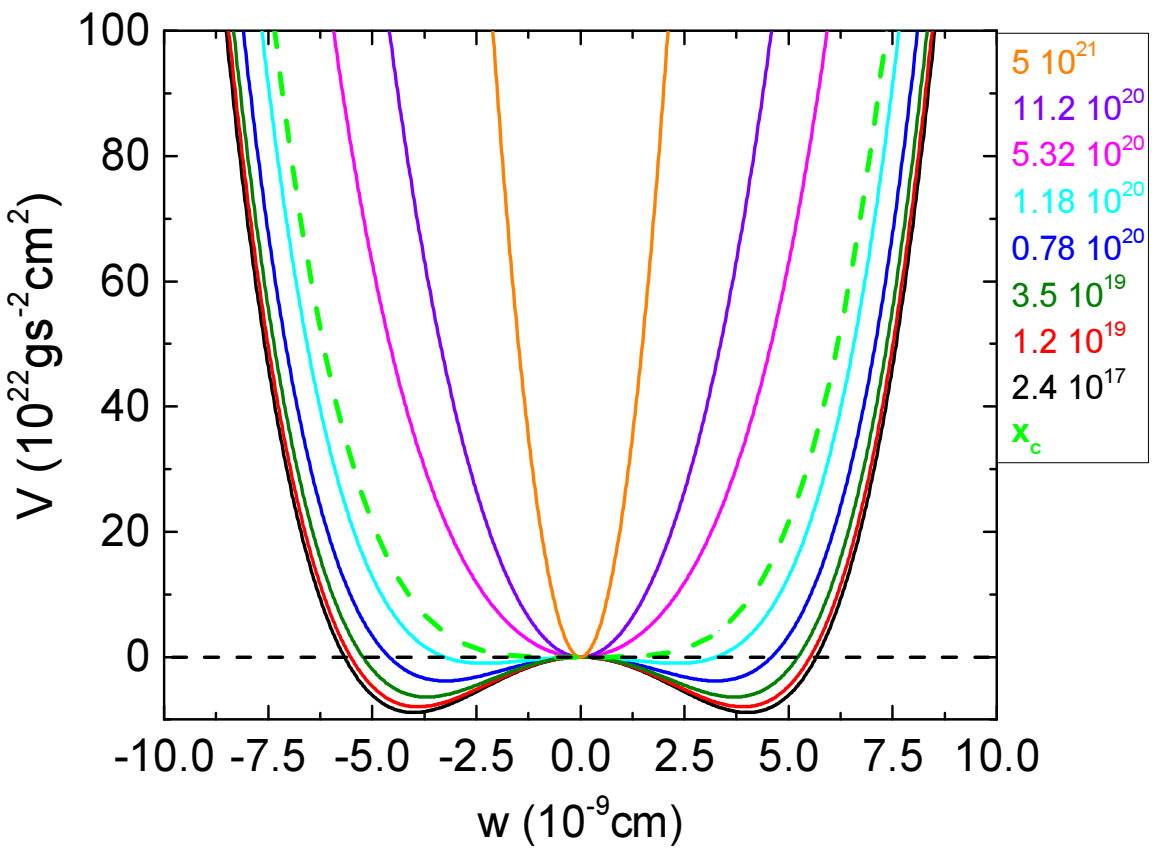

**Figure 2** Local potential $V$ in the core-shell related relative displacement coordinate $w$ for various carrier concentration $n$ ($cm^{-3}$) (the color code is given in the inset). The bright dashed green line corresponds to $n_c$.

Besides the fact that the potential changes from double-well to single well, it becomes also increasingly narrower with increasing $n$. This finding implies – counterintuitively – that the electrons within the $TiO_6$ unit are more strongly localized, i.e. $w \rightarrow 0$ see above, and dynamical p-d hybridization effects loose importance. Since this happens with increasing carrier concentration other pathways for the conductivity of doped STO must be present. In various recent work, extended studies of STO and other perovskites have shown that dislocations and defects form a kind of filaments (within the polar matrix) which are responsible for the conductivity [40 – 43].

The carrier concentration dependent values of $g_2$ and $g_4$ allow to calculate the temperature dependence of the lowest transverse optic and acoustic modes within the SPA. The squared optic mode frequency $\omega_{TO}^2(q = 0)$ is inversely proportional to the permittivity, whereas the acoustic mode experiences anharmonic mode-mode coupling at finite momentum as evidenced by anomalies in its dispersion [53 – 55]. This squared "soft" mode frequencies of doped STO are shown as a function of temperature in Figure 3.

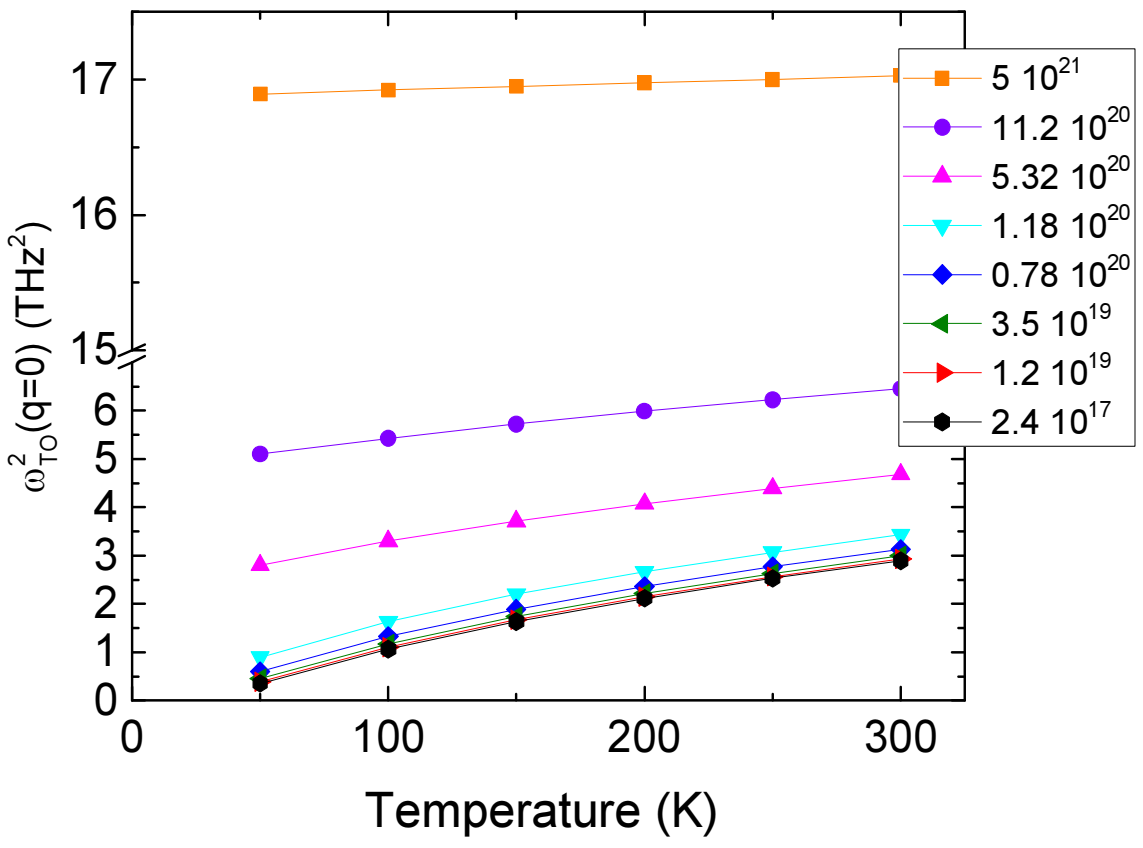

**Figure 3** The squared ferroelectric soft mode $\omega_{TO}^2$ as a function of temperature for various carrier concentrations $n$ ($cm^{-3}$) as given in the inset.

As is obvious from the Figure, they can be grouped into two categories, namely ferroelectric type up to $n_c$, and non-ferroelectric like for $n>n_c$ where only a minor softening with decreasing temperature is observed for $\omega_{TO}^2(q=0)$, consistent with the change in the potential from double to single well.

Since in undoped STO precursor dynamics have been predicted and experimentally confirmed [53 – 55], also the doped systems are investigated for these features which are signatures of polar or elastic nano domain formation. This is better carried out by inspection of the dispersion of the lowest optic and acoustic modes. Typical dispersions are shown for three carrier concentrations, low, intermediate and large, in Figures 4 a – c.

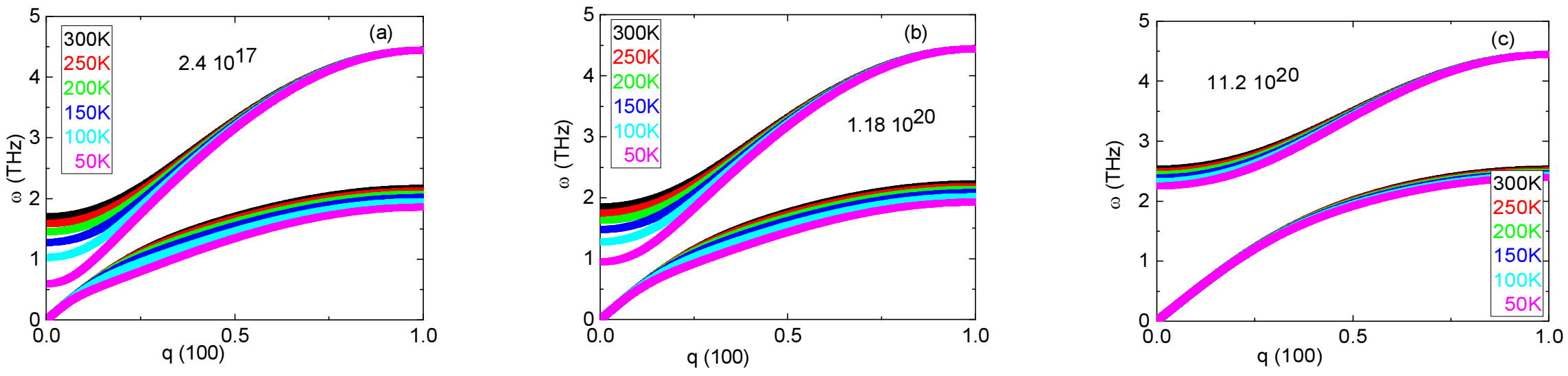

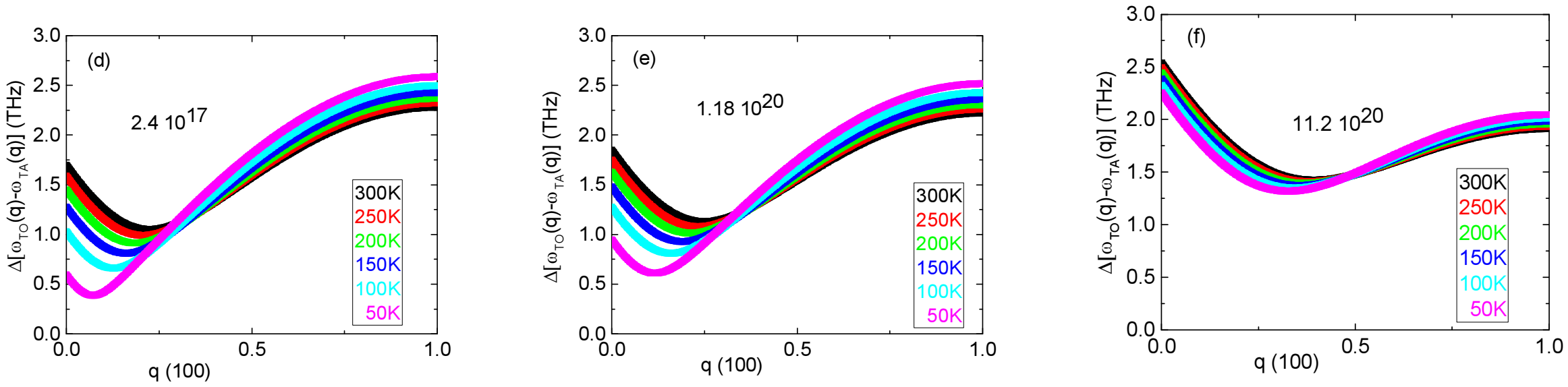

**Figure 4** Dispersion of the lowest transverse optic and acoustic modes for $n$=2.4×10$^{17}$cm$^{-3}$ (a), 0.78×10$^{20}$cm$^{-3}$ (b), 11.2×10$^{20}$ cm$^{-3}$ (c). The q-dependent frequency difference $\Delta[\omega_{TO}(q) - \omega_{TA}(q)]$ for $n$=2.4×10$^{17}$ cm$^{-3}$ (d), 0.78×10$^{20}$ cm$^{-3}$ (e), 11.2×10$^{20}$ cm$^{-3}$ (f). All curves are shown for different temperatures as given in the insets.

In the cases with small and intermediate carrier concentrations the increasing mode-mode coupling with decreasing temperature is apparent through anomalies at small momentum in the acoustic mode dispersion observed also experimentally [45, 46]. This has completely vanished for $n$=11.2×10$^{20}$ cm$^{-3}$ where the acoustic mode temperature dependence is almost absent especially in the small q range. The momentum values where the coupling is largest are clearly obtained by calculating the difference of optic and acoustic mode where the minimum indicates the spatial range of maximum mode-mode coupling. Again the three above examples are shown in Figure 4 d – f and all other cases lie intermediate or beyond those. With increasing carrier density the difference spectrum moves to higher energy and the minimum momentum value $q_{\min}$ shifts to higher q-values. This implies that the spatial extent of these nano domains decreases with increasing $n$ and its volume fraction shrinks. The related spatial spread of the polar region with radius $r_c$ as a function of temperature is shown for all investigated carrier concentrations in Figure 5. For carrier concentrations smaller than $n_c$, $r_c$ diverges with decreasing temperature and is only slightly dependent on the carrier concentration at high temperature to saturate at a spatial spread of approximately 5 lattice constants. This means that at all temperatures the system is inhomogeneous and consists of polar nano domains as matrix that coexists with conducting filaments. For $n$>$n_c$ small nano domains remain but their spatial extent and volume fraction are small and almost independent on temperature. In this case the sample inhomogeneity has almost completely vanished and the conductivity should be metallic-like as observed experimentally. It is important to mention that effective Hamiltonian approaches are unable to capture these “local” spatially confined features since they rely on the lattice periodicity. This is, however, absent in the present approach and appropriate modelling not given.

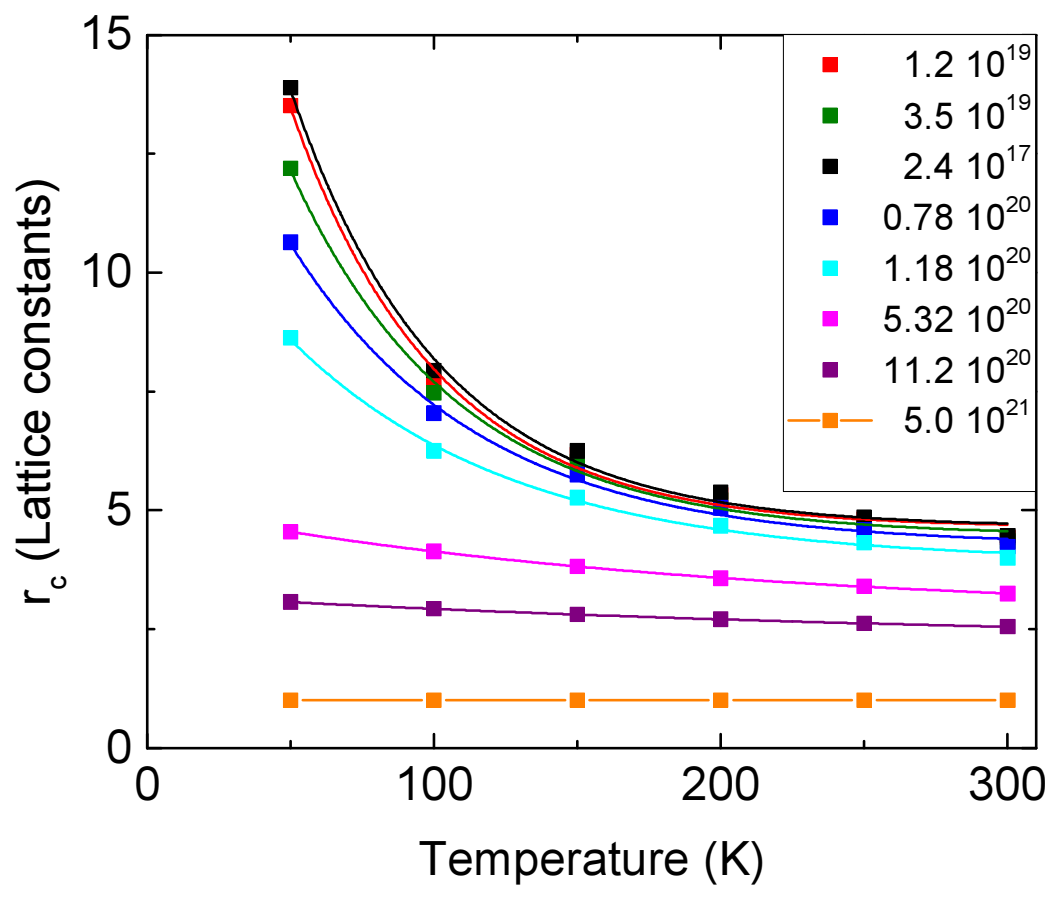

**Figure 5** Spatial extent of polar nano-regions $r_c$ as a function of temperature and various carrier concentrations $n$ as indicated in the inset to the figure.

All curves follow an exponential law: $r_c = r_0 + A\exp(-\frac{T}{T_K})$ where $r_0$ is a polaronic radius and of the order of several lattice constant for $n<n_c$ and decreases substantially for larger $n$. Similarly, $A$ is almost constant for small carrier concentrations and decreases rapidly for larger ones. The crossover temperature $T_K$ shows similar trends, but increases enormously for $n>n_c$. All parameters are summarized in Table 1.

**Table 1** Carrier concentration $n$, harmonic versus anharmonic coupling constant $-g_2/g_4$, polaronic radius $r_0$, exponential prefactor $A$, and crossover temperature $T_K$.

| Carrier concentration $n$ [$cm^{-3}$] | $-g_2/g_4$ [$10^{-18}cm^2$] | $r_0$ [lattice constants] | $A$ [lattice constants] | $T_K$ [K] |
|---|---|---|---|---|
| $2.4\ 10^{17}$ | 32.14 | 4.64 | 23.85 | 52.6 |
| $1.2\ 10^{19}$ | 30.46 | 4.63 | 23.34 | 52.6 |
| $3.5\ 10^{19}$ | 27.29 | 4.46 | 18.23 | 59.0 |
| $0.78\ 10^{20}$ | 21.08 | 4.26 | 13.57 | 66.7 |
| $1.18\ 10^{20}$ | 10.63 | 3.91 | 8.90 | 77.0 |
| $5.32\ 10^{20}$ | -52.13 | 2.80 | 2.29 | 185.0 |
| $11.2\ 10^{20}$ | -118.16 | 2.18 | 1.06 | 286.0 |

The question remains whether the above dynamical properties can be related to the metal/insulator (M/I) transition and superconductivity. This is certainly not possible directly, however upon multiplying the double-well potential depth with the density of states at $E_F$ and plotting this as a function of $n$ an approximate relation is obtained as displayed in Figure 6. As soon as the potential changes shape at $n_c$ the coupling is repulsive and not shown in the Figure.

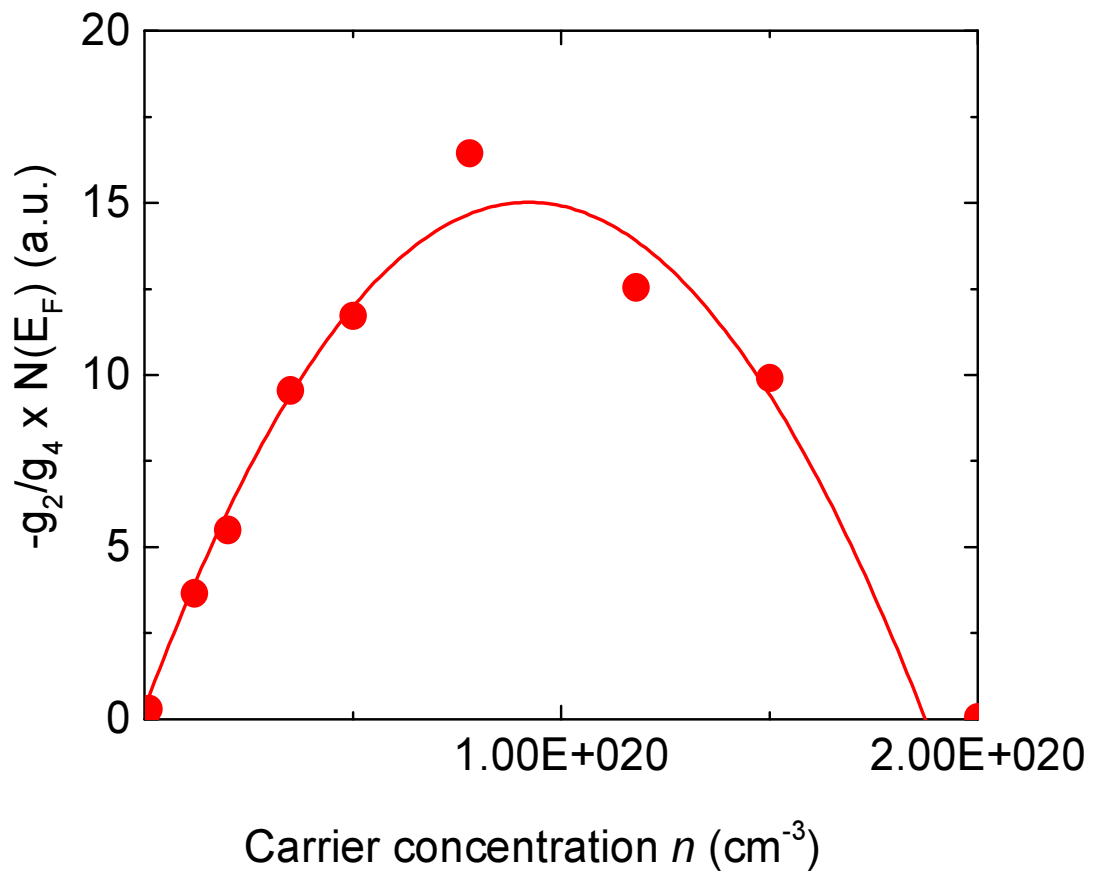

**Figure 6** Effective coupling constant $-g_2/g_4$ as a function of carrier concentration $n$. Calculated results correspond to full circles, line is a parabolic fit.

Obviously a dome-like dependence of the "effective" coupling constant on $n$ is obtained which mimics the dependence of $T_c$ on $n$ as observed experimentally [21]. It is, however, within the above presented results, not possible to link directly Figure 6 with superconductivity. Hence an Eliashberg type calculation is in progress, and we are convinced that the data will indeed be relevant to superconductivity in STO. If so, it must have been concluded that superconductivity occurs in a region with strong inhomogeneity where filamentary conductivity coexists with a polar matrix. It is worth mentioning that similar conclusions have already been reached for cuprate superconductors [56].

## *Experimental results*

The analysis of the electrical properties of doped STO is based on the volume concentration of carriers. Self-doped STO samples exhibit different electronic, chemical, and structural properties as compared to the matrix which is little affected by doping [57]. However, in the core dislocations the removal of oxygen is preferentially taking place along extended defects [40, 58]. Consequently, a *local* reduction is observed, which cannot be measured by average macroscopic property detecting tools like, e.g., mobility and Hall measurements. This statement is supported by local conductivity atomic force microscopy (LC-AFM) data [39, 40], which evidences that very low concentrations of oxygen deficiency assemble along the dislocations. In order to fulfill the Mott criterion for the M/I transition in reduced STO these conclusions are necessary and sufficient. From a LC-AFM analysis of *in-situ* and *ex-situ* cleaved surfaces of reduced STO and the in-plane cleaved crystals the resistivity data (Table 2) have been compared with the four point method taken at the maximum reduction process [e.g. 60]. The measured resistance using 4-point method calculated resistivity values of Table 2 are in close agreement with data from Spinelli [36] and Schooley [18] for reduced STO either being metallic or superconducting. Since the reduction methods of [18, 36] are not commonly used and differ from the ones of [40, 58] we cannot prove that either the M/I transition or the superconducting one are confined to core dislocations or filaments. In order to provide more stringent evidence for our arguments, which do not support a statistical distribution of “defects”, a simple but impressive experiment has been carried out, which highlights the important role of dislocations and filament formation in the electric transport of reduced $SrTiO_3$.

**Table 2** Resistivity and global and local concentration of oxygen defects induced in $SrTiO_3$ single crystal at different reduction temperatures.

| Reduction temperature (°C) | Single crystal resistance assigned by four points method (Ω) | Calculated volume resistivity (Ωcm) | Resistivity (Ωcm) at room temperature for vacuum reduced $SrTiO_3$, after Spinelli et al [36] | Hall concentration of carriers N at at room temperature ($N/cm^3$), after Spinelli et al. [36] | Calculated volume concentration of oxygen vacancies N due to oxygen effusion per unit volume ($N/cm^3$) [38] | Effective oxygen vacancies non-stoichiometry x in $SrTiO_{3-x}$ derivate from [38] | Average concentration of oxygen vacancies N in the core of dislocations network ($N/cm^3$) determined from effusion data and extension of the hierarchic tree of dislocations [38] |
|---|---|---|---|---|---|---|---|
| 600 | 1320 | 49 | 18.2 | $5.64\ 10^{16}$ | $2.2\ 10^{13}$ | $4.4\ 10^{-10}$ | $3.3\ 10^{19}$ |
| 620 | | | $2.2\ 10^{2}$ | $1.04\ 10^{16}$ | | | |
| 650 | | | $89.1\text{-}2.2\ 10^{3}$ | $3.8\ 10^{15}\text{-}1.98\ 10^{16}$ | | | |
| 700 | 118 | 4.43 | $3.87\text{-}1.13\ 10^{3}$ | $3.76\ 10^{15}\text{-}2.13\ 10^{17}$ | $3.4\ 10^{13}$ | $1.1\ 10^{-10}$ | $8.4\ 10^{19}$ |
| 800 | 47<br>150 [61] | 1.76<br>4.69 | | | $2.6\ 10^{13}$<br>$3\ 10^{14}$ [37] | $1.75\ 10^{-9}$ | $1.3\ 10^{20}$<br>$2\ 10^{20}$ [37] |
| 900 | 27<br>25 | 1.01<br>0.78 | | | $9.1\ 10^{13}$ | $3{,}34\ 10^{-9}$ | $2.6\ 10^{20}$ |
| 1000 | 14.8 | 0.56 | | | $8.0\ 10^{13}$ | $5.03\ 10^{-9}$ | $3.8\ 10^{20}$ |
| 1100 | | | 0.15-0.47 | $2.23\ 10^{18}$ | | | |

A reduced STO single crystal is cut into two identical parts where one has the original epi-polished surface whereas the other is scratched by using e.g. a diamond pyramid (Figure 7).

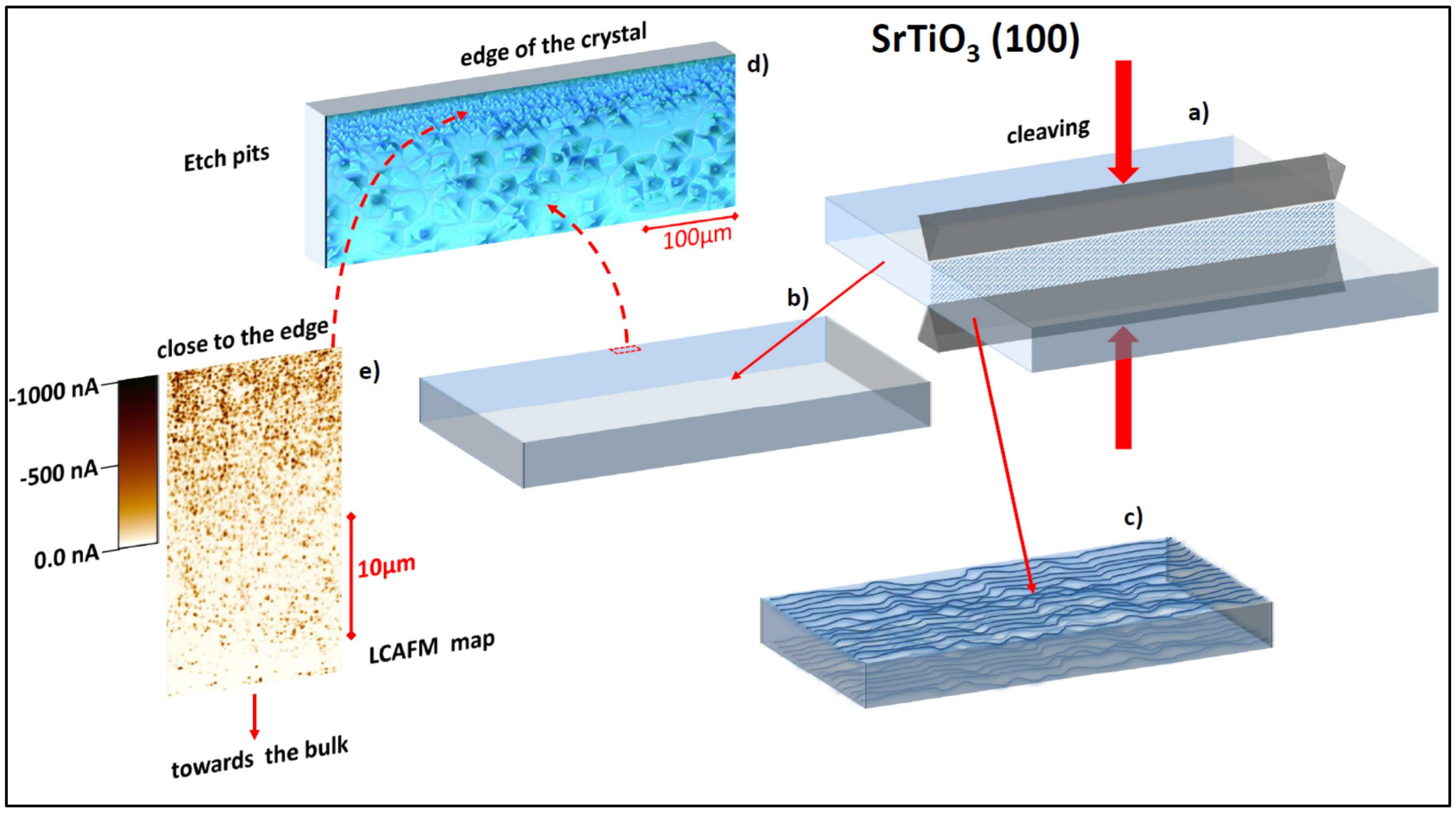

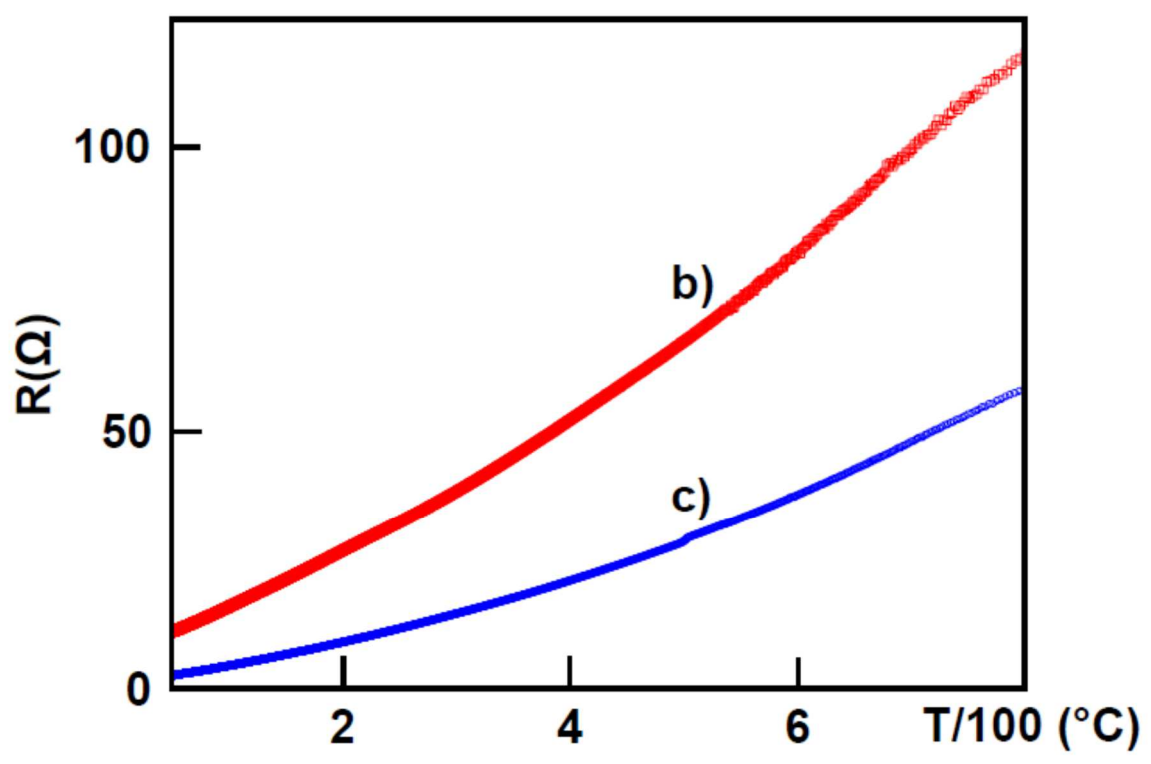

**Figure 7** (Top) Schematic view of preparation of STO sample with different dislocation densities in the surface layer. After cleaving the sample a) the distribution of dislocations and their concentration is identical to the reference crystal (~$10^{9\text{-}10}$/cm$^2$) in contrast to c) where the surface area has been scratched and thereby has generated $10^{11\text{-}12}$ dislocation/cm$^2$. The upper inset d) evidences that the edge of the reference crystal contains a substantially larger dislocation density as compared to b). The LC-AFM mapping e) of the part of the crystal close to the edge of the reduced crystal shows an extremely high concentration increase of the conducting dislocations. The ratio of the conductivity of dislocations (represented by dark-brown color) to the matrix conductivity (represented by white color) is of about 3-4 orders of magnitude. (Bottom) Temperature dependence of the resistivity of samples b and c. Sample b is represented by the red line, and c by the blue one.

This procedure has increased the number of dislocations by one to two orders of magnitude from $10^{9\text{-}10}$/cm$^2$ to $10^{11\text{-}12}$/cm$^2$ of the scratched sample [59]. Upon reducing both samples simultaneously under the same thermodynamic conditions, the resistivity of both has been monitored as a function of temperature (figure 7b bottom). As is obvious from the figure (7b bottom), the electrical transport properties of both samples are very different from each other in spite of the use of *identical reduction processes*. Even though both samples show metallic behavior, the resistivity of the polished one is shifted to higher values as compared to the scratched where not just a scaling factor enters. Apparently, the increase in the dislocation density of the scratched sample (c) by two orders of magnitude as compared to the epi-polished one (b) has decreased the resistivity by a factor of five, which intuitively was to be expected. While this observation seems to be counteracting the initial argument, a decisive fact has been omitted, namely, the dislocation density on the edges (upper inset to Fig. 7 b top) of both specimen are alike. By taking this fact into account, a simple count of dislocations stemming from both, the faces and the edges of both samples contributing to the resistivity, cannot be larger than approximately five as deduced from detailed resistivity measurements including the edges [38,]. The role of dislocations played for the electrical transport can indirectly be proven through measurements of the temperature gradient (using an infrared camera) along the sample during poling. In the vicinity of regions with a high concentration of these extended defects the local temperature is increased by 10 to 40K caused by the preferential current flow and the related Joule loss [59]. Thus this experiment supports the basic assumption of filamentary conductivity in doped STO which is intimately related to the two component approach suggested to be realized in this system, an intact dielectric matrix and a hierarchic 3D network of metallic/superconducting filaments. An important ingredient is attributed to the interfaces between both which mediates their interplay and most likely acts as an inter-band interaction term analogous to two-band superconductivity. A schematic view of the above scenario is shown in Figure 8 where doping is shown to be constrained to the core of filaments, whereas polar properties are limited to the remaining regions rendering the system intrinsically inhomogeneous.

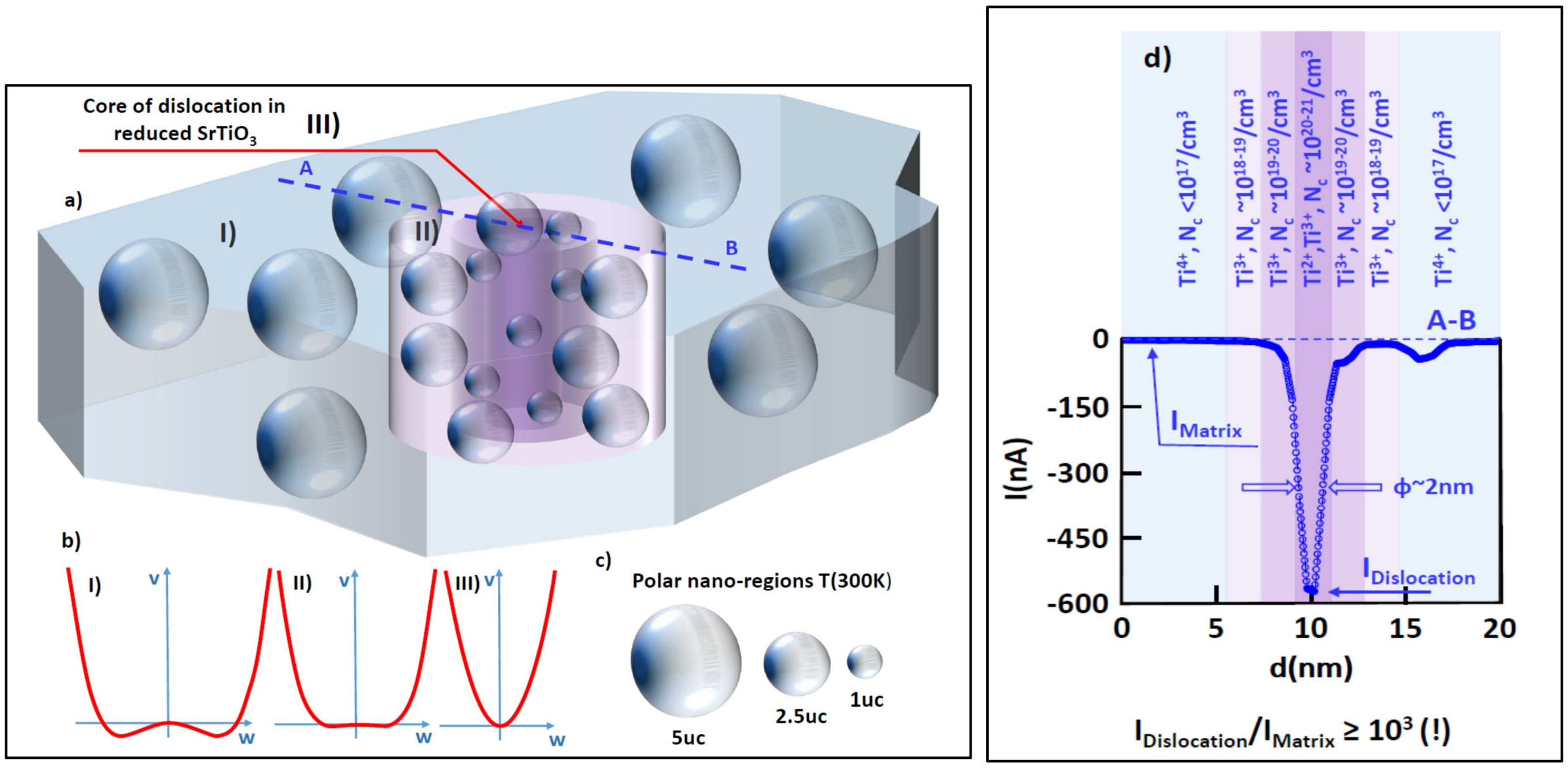

**Figure 8** Schematic representation of the distribution of polar nano clusters in reduced STO (insets a, and the corresponding local potential b-I), close to the conductive core of the dislocation (insets a, with the potential b-II) and inside the core (insets a, and potential b-III). In the matrix, far from the dislocations, the concentration of carriers is rather small and thus admits its unhindered clustering (inset b-I). For details see Figure 2 and 5 with a relatively large spatial extent (here ≈5 lattice constants (uc) at 300K (inset c)). In the vicinity of conductive dislocations their expansion is reduced to 2.5uc (insets a, c) caused by the increase in the local concentration of d-electrons up to $10^{17\text{-}19}/cm^3$ (inset d, cross-section A-B, and inset II-b). In the core of the dislocation, the concentration of carriers reaches a maximum value of $10^{20\text{-}21}/cm^3$ which renders the formation of nano clusters unlikely (insets b-III and c). In the core of the dislocation with a typical radius of 2nm, their size is not larger than 1 uc (insets a and c). Note, the average concentration of Ti d-electrons has been calculated from effusion data [37, 38], and the change in carrier concentration, which depends on the distance from the core, was derived from the LC-AFM measurements as depicted in the inset d (for details see e.g. [38 – 40]).

As outlined above, even with increasing carrier density, mode softening persists on a dynamic time scale of ps and length scale of nm. This process can be confirmed experimentally on the nano-scale by using, e.g. time-resolved infrared spectroscopy or scanning near field optical microscopy (SNOM). Alternatively, piezoelectric force microscopy (PFM) offers the possibility to detect locally a piezoelectric response which is an indirect probe of mode softening through the creation of induced dipole moments. This has been shown in [60, 61] where in the vicinity of the core of edge dislocations in a STO bicrystal a polarization of the order of 20μC/cm$^2$ has been detected. Analogously, at steps of a terrace piezoelectric activity can be expected [62, 63]. This is demonstrated in Figs. 9 where along a sharp step of a plastically deformed STO crystal, corresponding to a broad band of dislocations, piezoelectric responses are observed.

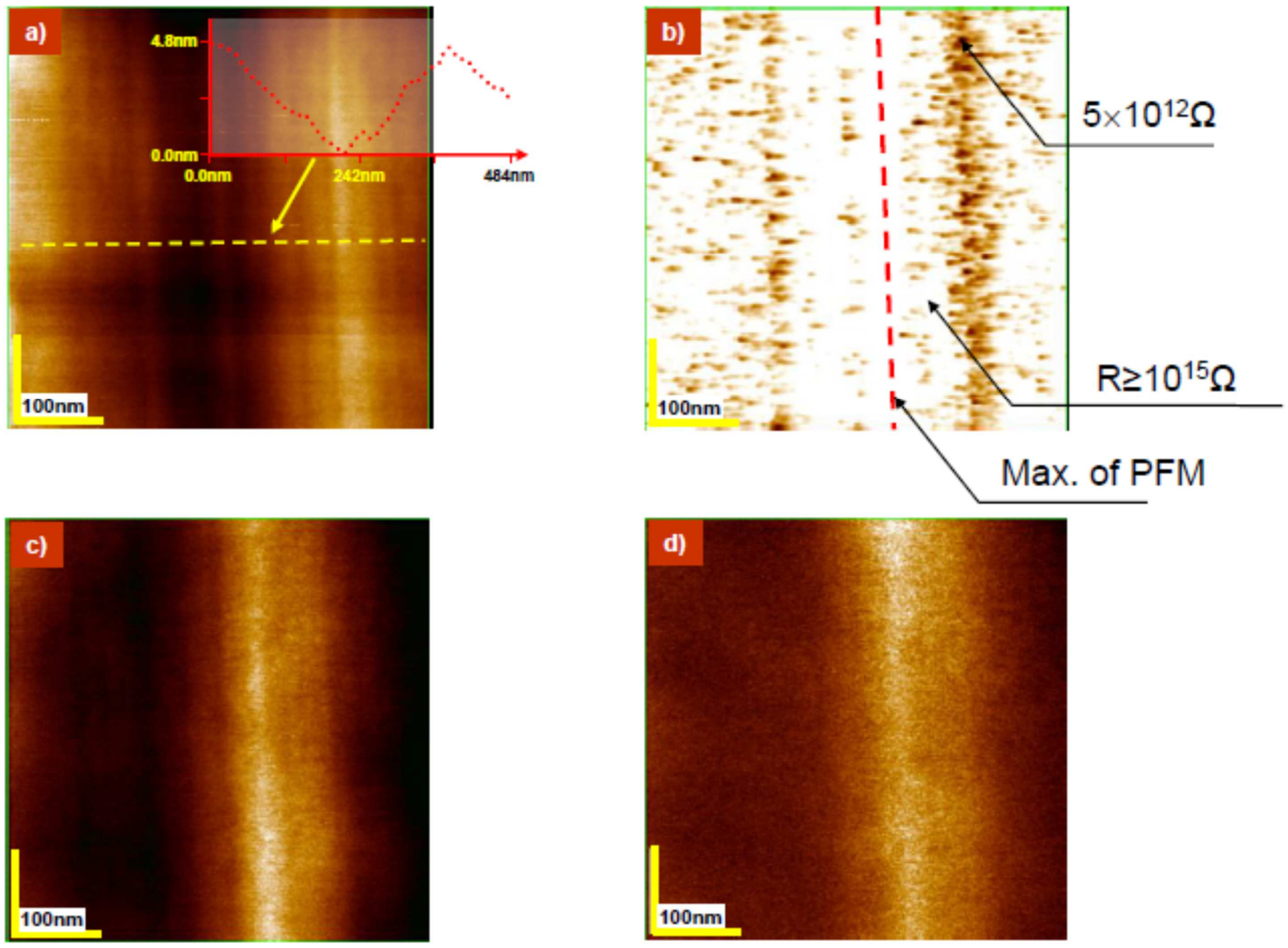

**Figure 9** Topographic image of a step (AFM) of a plastically deformed stoichiometric $SrTiO_3$ crystal (a). Near the step there is a wide band of conductive dislocations/filaments (dark points on LCAFM-map (b). The position of the filaments is correlated with the position of the etch pits (see [38]). Along the step, a high piezoresponse (out-of-plane c), and in-plane d)) with clear maximum of the PFM response, can be observed between both rows of conductive dislocations. Note: The rest of the crystal far from the step shows very low piezo activity.

Using highly sensitive LCAFM mapping of the near-step area, a series of dislocations (filaments), with considerably lower resistance than the rest of the stoichiometric crystal, exists (Fig. 9b). This suggests that in the core of the dislocations the local stoichiometry is different from the one of the bulk.

When a crystal with semiconducting dislocation cores, dislocation bundles, steps or with low/high angle boundary is additionally reduced, the dislocations are transformed into metallic filaments. Obviously, these affect not only the electrical properties (as, e.g., the insulator-metal transition), but produce also enormous changes in the dielectric and piezoelectric properties. In the areas of high (metallic) conductivity the piezoelectricity vanishes (Fig. 10). This finding clearly questions the validity of the concept of metallic-ferroelectricity [64], or more precisely metallic-piezoelectricity. On the contrary, and considering the above results, the concept of essential heterogeneity in thermally reduced $SrTiO_3$ crystals, namely a dielectric (ferroelectric) matrix and 3D network of metallic filaments, provides a reasonable and realistic picture of perovskite oxides.

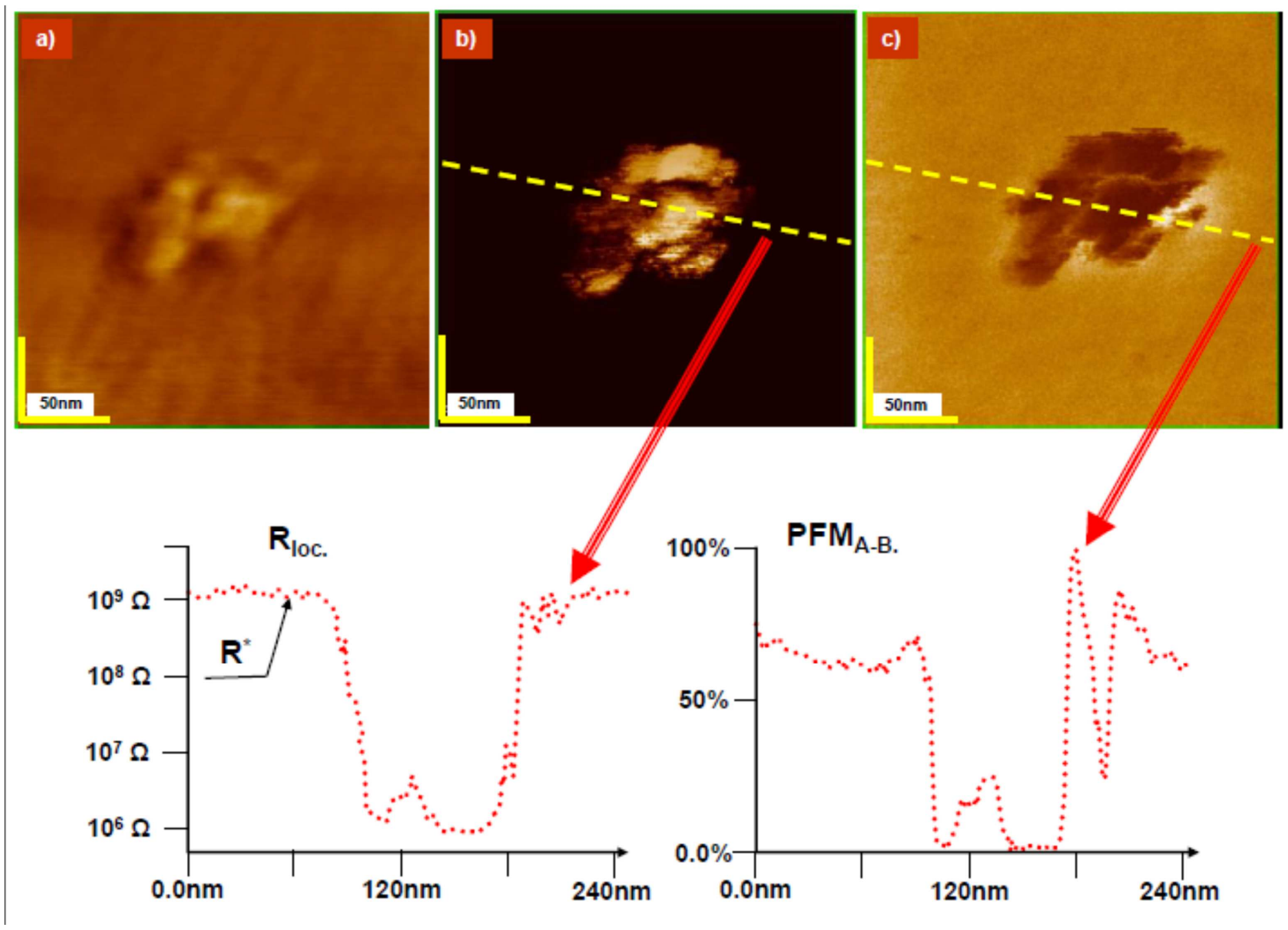

**Figure 10** Topographic image of the exit of a dislocation bundle in thermally reduced STO: (a) the topography variation is about 0 – 4.8nm. In the area of the bundle (LCAFM) (b) with metallic properties the piezoresponse (c) is absent. This can be observed at the cross sections of LCAFM (b) and PFM (c) responses. The distribution of the resistance along the cross section (lower part of Fig. 10) shows an anti-correlation with the distribution of the piezo-activity on the cross section of the PFM. The resistance outside the bundle is much higher than shown on the cross section, which is related to the finite resolution of the A/D converter (here 4 decades), so that using additional measurements (only in this area) the resistance is higher than $10^{12-13}$ Ω.

It is important to note that aliovalent doping of STO (with, e.g., La or Nb), causing n type conductivity and an insulator metal transition for only 0.1-1% doping, leads to a more homogeneous distribution of carriers in the matrix. This implies that in STO:La or STO:Nb the formation of polar nano-regions with the same radii throughout the crystal volume takes place, contrary to thermally reduced STO crystal shown in Figure 8.

### *Conclusions*

In conclusion, the dynamical properties of STO have been calculated as a function of carrier concentration. Up to a critical concentration $n_c$ the lattice potential is of double-well character and strong transverse optic mode softening takes place. This is linked to the formation of polar nano regions which grow in size with decreasing temperature implying substantial sample inhomogeneity. The "insulating" nano domains coexist with the filamentary conductivity, and hence possible links to superconductivity have been suggested. As long as this coexistence persists, superconductivity survives. It suggests that the essential origin of it must be associated with the polar character and the two-component properties. Beyond $n_c$ almost "normal" dynamics are observed with negligible mode softening and nano domain formation. Typical metallicity is expected there and homogeneous sample properties.

The theoretical results are supported by experiments which demonstrate the local metallic character and the insulating behavior of the matrix. From additional PFM and LCAFM data, the dielectric character of the matrix has been confirmed together with the strong variations in the resistance upon crossing from the matrix to the metallic filaments. Here, especially, the comparison of stoichiometric crystals with scratched (i.e. with high concentration of mechanically generated dislocations), and other epi-polished ones (with a low concentration of dislocations) has provided essential clues in understanding the role of filaments for the electrical transport phenomena (in particular for the analysis of the type of transition I/M induced by reduction under vacuum conditions in the network of dislocation filaments), and in modifications of local polarization in the vicinity of conductive core of dislocations (filaments).

Authors' contribution statement

A.B.-H. performed the theoretical calculations. K.S. carried the experiments through. H.K., K.R. and A.S. contributed through numerous discussions. All authors reviewed the article.

The corresponding author is responsible for submitting a competing interests statement on behalf of all authors of the paper.

Competing interests

The authors declare no competing financial AND non-financial interest.